\documentclass[reprint,superscriptaddress,prl,longbibliography]{revtex4-2}
\usepackage[utf8]{inputenc}
\usepackage[T1]{fontenc}
\usepackage{amsmath}
\usepackage{amssymb}
\usepackage{mathrsfs}
\usepackage{graphicx}
\usepackage[capitalize, nameinlink]{cleveref}
\usepackage{subfigure}
\graphicspath{{figures/}}
\usepackage{adjustbox}
\usepackage{booktabs}
\usepackage{multirow}

\usepackage{color}
\usepackage{rotating}
\usepackage{ulem}
\usepackage{bm}

\begin{document}

\title{Non-equilibrium formulation of helicity-dependent thermal field for ultrafast magnetization dynamics}

\author{Ezio Iacocca}
\affiliation{Center for Magnetism and Magnetic Nanostructures, University of Colorado Colorado Springs, Colorado Springs, CO 80918, USA}

\date{\today}

\begin{abstract}
Far-from-equilibrium magnetization dynamics can be accessed when a magnetic material is subject to a femtosecond excitation, such as an optical laser or an electric current. Numerically, the demagnetization of magnetic materials is typically modeled by atomistic spin dynamics. Micromagnetic models generally fail to reproduce ultrafast demagnetization in a grid independent manner. Here, we propose a non-equilibrium thermal field whose features depend on atomic spin flip probabilities. Under the assumption that each spin flip is equivalent to a quantum of angular momentum, equivalent temperatures on the order of thousands of Kelvin are achieved. Demagnetization is quantitatively reproduced for several cell sizes. The presented approach can be further refined and extended towards a grid-independent and multiscale modeling of ultrafast magnetization dynamics.
\end{abstract}

\maketitle


The effect of temperature on the magnetization of a single-domain particle can be expressed as a thermal field~\cite{Brown1963} with a magnitude that obeys the fluctuation-dissipation theorem~\cite{dAquino2006}. This approach is strictly valid when the particle is at thermal equilibrium. The field magnitude is proportional to $V^{-1/2}$, where $V$ is the particle's volume, indicating that the magnetization of a large magnetic volume fluctuates less, on average. Within a micromagnetic framework~\cite{Brown1963b}, where a magnetic particle is discretized in cells, the thermal field is applied on each cell. Each of these cells has dimensions on the order of the exchange length, implying that the fluctuations are large compared to the total particle's volume but the average remains low due to the uncorrelated field. This approach has been extensively used in the research of spin-torque nano-oscillators, see e.g. Refs.~\cite{Xiao2005,Thadani2008,Carpentieri2010,dAquino2011b,Dumas2013,Puliafito2019}.

Even though the thermal field was derived in thermal equilibrium, it has been applied to model ultrafast magnetism~\cite{Kirilyuk2010} in atomistic spin dynamics~\cite{Eriksson2017}, returning excellent agreement with experiments~\cite{Radu2011,Ostler2012,Iacocca2019}. In fact, it has been shown that purely thermal effects can drive ultrafast magnetization dynamics by optical lasers, e.g.,~\cite{Ostler2012,Mangin2014}, and even electrical pulses~\cite{Yang2017,Wilson2017,Jhuria2020}. However, these ultrafast effects are generally not possible to reproduce quantitatively with micromagnetic simulations where the cells are composed by hundreds of atoms. This imposes a tremendous limitation in the modeling of ultrafast magnetization dynamics that require a full micromagnetic approach, e.g., domain patterns~\cite{Jangid2023} and solitons~\cite{Turenne2022}, as well as the full profile of optical lasers. It must be noted that descriptions based on a temperature-dependent dynamic equation, the Landau-Lifshitz-Bloch equation~\cite{Evans2012,Atxitia2010,Jakobs2022}, have been extended to a micromagnetic description with excellent agreement with ASD in the case of ferrimagnets~\cite{Raposo2022}.

In this letter, we derive a non-equilibrium contribution to the thermal field that is grid-independent and suitable for ultrafast magnetization dynamic simulations. The non-equilibrium field has two main contributions. First, the magnitude of the field includes the energy contribution from spin flips, proportional to the number of spins and their spin flip probability. Second, the distribution of the thermal field is determined by the probability distribution of spin flips. In contrast to the uniformly distributed thermal field expected from uncorrelated noise, the derived contribution has a mean and standard deviation, converging to a Gaussian for a large number of spins. A key advance of our work is the quantification of the magnon energy as a function of cell-size, which allows for a consistent and grid-independent determination of the non-equilibrium field magnitude.

Our starting point is a statistical model for spin flips. This approach was introduced in Ref.~\cite{Gorchon2016} to model multishot switching in ferromagnets. In this spirit, we consider a left circularly polarized laser pulse, with helicity $\sigma$. This laser favors a down spin orientation which is modeled in Ref.~\cite{Gorchon2016} as a differential temperature between up ($T_\uparrow$) and down ($T_\downarrow$) spins,
with $T_\uparrow>T_\downarrow$. Here, we will consider the probabilities of switching for each spin
\begin{equation}
    \label{eq:probs}
    \begin{cases}P\quad\text{if spin-up}\\B\quad\text{if spin-down}\end{cases}.
\end{equation}

For our selected helicity, $P>B$ meaning that it is more likely for an up spin to switch due to the optical laser. In this framework, a small probability implies that the laser fluence is low, while $P=B$ implies linear polarization.

Let us consider a micromagnetic cell with $N$ spins. We represent the number of up spins as $u$ and the number of down spins as $d$. Therefore, $N=u+d$. The normalized magnetization of the cell is given by $m_z=(u-d)/N$, where we set the $z$ direction as our spin basis. In other words, we assume the material has perpendicular magnetic anisotropy. This definition is simply a quantization of the $m_z$ component that limits to a continuum as $N\rightarrow\infty$. The pertinent question is, given an initial $m_z^i$, what is the probability for $m_z^f$ to occur? To solve this problem, we consider a number of switching events, $n$ for the up spins and $m$ for the down spins. Therefore, the final magnetization is
\begin{equation}
    \label{eq:mzfinal}
    m_z^f = \frac{U-D}{N}=\frac{2U-N}{N},
\end{equation}
where $U=u-n+m$ and $D=d+n-m$ are the final up and down spins, respectively. The probability of achieving $m_z^f$ from $m_z^i$ is given by the sum of probabilities $P_{ud,nm}$ for all the available combinations of $n$ and $m$ that achieve $U$ and $D$ given $u$ and $d$,
\begin{equation}
    \label{eq:probnm}
    P_{ud,nm} = \frac{u!d!P^n(1-P)^{(u-n)}B^m(1-B)^{(d-m)}}{n!m!(u-n)!(d-m)!}.
\end{equation}

For example, given $N=4$, the initial magnetization $m_z^i=0.5$ is obtained by any permutation of the spin arrangement $\uparrow\uparrow\uparrow\downarrow$. A transition to $m_z^f=m_z^i$ is possible if $n=m=0$ (no switching events) or if $n=m=1$, resulting in the probability $P_{31,00}+P_{31,11}$.

In the limit of large $N$, the central limit theorem ensures that the distribution of $m_z^f$ will be Gaussian. Therefore, we can define its mean and standard deviation:
\begin{subequations}
\label{eq:meanstd}
\begin{eqnarray}
    \label{eq:mean}
    \mu_{PB} &=& \sum_{U,D}\frac{U-D}{N}P_{ud,nm},\\
    \label{eq:std}
    \sigma_{PB}^2 &=& \left[\sum_{U,D}\left(\frac{U-D}{N}\right)^2P_{ud,nm}\right]-\mu_{PB}^2,
\end{eqnarray}
\end{subequations}
where the sums are performed over all $U,D$ given an initial $u$ and $d$ subject to switching events $n$ and $m$.

We now connect this description with that of a magnetic field. The main assumption in this letter is that the distribution quantified by Eqs.~\eqref{eq:meanstd} determines the distribution of a non-equilibrium field. We define the non-equilibrium field as $\mathbf{H}_{n-e}=H_{n_e}\mathbf{\hat{h}}_{n-e}$. The unit vector is given by
\begin{subequations}
\label{eq:Hne}
\begin{eqnarray}
    \label{eq:Hnex}
    \mathbf{\hat{h}}_{n-e}\cdot\hat{\mathbf{x}} &=& \arcsin{\eta}\cos{\phi},\\
    \label{eq:Hney}
    \mathbf{\hat{h}}_{n-e}\cdot\hat{\mathbf{y}} &=& \arcsin{\eta}\sin{\phi},\\
    \label{eq:Hnez}
    \mathbf{\hat{h}}_{n-e}\cdot\hat{\mathbf{z}} &=& \eta,
\end{eqnarray}
\end{subequations}
where $\eta$ is a random number obeying a Gaussian distribution with mean $\mu_{PB}$ and standard deviation $\sigma_{PB}$ while $\phi$ is a uniformly distributed random number in the range $[0,2\pi]$. This is the main difference with the traditional thermal field which is uniformly distributed on a sphere.

The magnitude of the non-equilibrium thermal field, $H_{n-e}$, is determined by the number of spin flips expected from the distribution. Given the initial magnetization $m_z^i$ with $u$ and $d$ spins, the final magnetization is simply $m_z^f=\mu_{PB}$, leading to expected $u'$ and $d'$ spins. Each flipped spin with respect to the background contributes to a unit of angular momentum $\hbar$. Therefore, the difference in magnon energy in a cell with $N$ spins during a time $\delta t$ is
\begin{equation}
    \label{eq:magnonEnergy}
    \Delta E = \left|\frac{N-|u-d|}{2}-\frac{N-|u'-d'|}{2}\right|\frac{\hbar}{\delta t},
\end{equation}
where the absolute values ensure that the maximum magnon energy occurs when the spin distribution is $u=d$ or $u'=d'$. This energy is large for small values of $\delta t$, which is important when considering the numerical integration scheme. Our assumption is that this energy can be considered to be an additional contribution to the thermal field with magnitude~\cite{dAquino2006}
\begin{equation}
    \label{eq:HneMag}
    H_{n-e} = \sqrt{\frac{2\alpha\Delta E}{\gamma\mu_0^2M_sV\delta t}}.
\end{equation}
For small cells and short times, this field can be substantial; on the order of $10^{12}$~A/m which corresponds to tens of Tesla. Equivalently, we can equate $\Delta E=k_BT$, where $k_B$ is Boltzmann's constant, to obtain a corresponding temperature on the order of $10,000$~K. In the atomic limit, $u,d$ can only take on 0 or 1 and $N=1$, therefore $\Delta E=0$ and the correction does not apply. Therefore, the correction is only meaningful for $N>1$, i.e., when cell sizes are larger than the atomic lattice.

The presented approach is verified numerically within a pseudospectral Landau-Lifshitz (PS-LL) model~\cite{Rockwell2024,Roxburgh2025b}. In this model, the traditional Landau-Lifshitz equation is used
\begin{equation}
    \label{eq:LL}
    \frac{\partial}{\partial t}\mathbf{m} = -\gamma\mu_0\left[\mathbf{m}\times\mathbf{H}_\mathrm{eff}+\alpha\mathbf{m}\left(\mathbf{m}\times\mathbf{H}_\mathrm{eff}\right)\right]
\end{equation}
where $\mathbf{m}$ is the magnetization vector normalized to the saturation magnetization $M_s$, $\gamma$ is the gyromagnetic ratio, $\mu_0$ is the vacuum permeability, $\alpha$ is the Gilbert damping constant (valid in this form if $\alpha\ll1$), and the effective field $\mathbf{H}_\mathrm{eff}$ is given by
\begin{equation}
    \label{eq:Heff}
    \mathbf{H}_\mathrm{eff} = \mathbf{H}_l+\mathcal{F}^{-1}\{\kappa(k)\mathbf{\hat{m}}\}+\mathbf{H}_{n-e}+\mathbf{H}_\mathrm{th},
\end{equation}
where $\mathbf{H}_l$ are local fields including anisotropy and external fields, $\mathcal{F}^{-1}\{\cdot\}$ represents the inverse Fourier transform, $\mathbf{\hat{m}}$ is the Fourier transform of the spatial magnetization, $\kappa(k)$ is a convolutional kernel that describes the nonlocal interactions, and $\mathbf{H}_\mathrm{th}$ is the traditional thermal field. In this paper, we consider only the exchange interaction, so $\kappa(k)=-M_s\omega(k)$ where $\omega(k)$ is the dimensionless magnon dispersion relation, {$\omega(k)=2\left(\frac{\lambda_\mathrm{ex}}{a}\right)^2\left[\sum_i1-\cos(ak_i)\right]$}, $a$ is the lattice constant, $\lambda_\mathrm{ex}=\sqrt{2A/(\mu_0M_s^2)}$ is the exchange length as a function of the exchange constant $A$, and the sum is over the spatial coordinates, $i=x,y,z$. The 2D implementation of the PS-LL has been used in the context of ultrafast magnetism in Ref.~\cite{Foglia2024}. Here, we use a 3D implementation of the PS-LL equation that follows closely the 2D implementation, and its details will be discussed elsewhere.

To verify the implementation of the proposed approach, we model the ultrafast demagnetization of an approximate thin film. We use a domain of $60$~nm~$\times60$~nm~$\times4$~nm with periodic boundary conditions in the plane and reflecting boundary conditions normal to the plane. This small volume is chosen with the intent to performing grid-dependent simulations. The non-local dipole field is neglected to ensure that the simulation behaves as a thin film and because it is typically not crucial at ultrafast timescales. Therefore, the local field $\mathbf{H}_l=(H_k-M_s)\hat{z}$ is the local anisotropy approximation for a thin film, where $H_k$ is the perpendicular magnetic anisotropy field.

As an example, we use material parameters of FePt~\cite{Gorchon2016}: saturation magnetization $M_s=950$~kA/m, uniaxial anisotropy field $H_k=7086$~kA/m, exchange constant $A=4.1$~pJ/m leading to an exchange length of $\lambda_\mathrm{ex}=2.69$~nm, and $\alpha=0.02$. The lattice constant is estimated to be in the range of $3d$ transition metal alloys, $a\approx0.4$~nm. With these parameters, simulations can be performed with cubic cells up to a side length $\lambda_\mathrm{ex}$, which leads to a rather large number of spins, $N=(\lambda_\mathrm{ex}/a)^3\approx304$. The calculation of the distribution's mean and standard deviation requires the computation of the probabilities $P_{nm}$, Eq.~\eqref{eq:probnm}, which features extremely large factorials. To circumvent this computational issue, we compute $P_{nm}=e^{\ln P_{nm}}$ when $N>100$, where $\ln{P_{nm}}$ can be solved using Stirling's formula for the logarithms of factorials. 

To model ultrafast demagnetization, we assume a spatially uniform femtosecond pulse with a Gaussian temporal profile. The cusp is centered at $t_c=2$~ps and the standard deviation is $t_\sigma=0.5$~ps leading to a pulse duration (full-width and half maximum) of approximately $1.17$~ps. The pulse is represented as an effective laser temperature $T(t)=T_0e^{-(t-t_c)^2/(2t_\sigma^2)}$.

\begin{figure}[t]
\includegraphics[width=3in, trim={0 0 0 0}, clip]{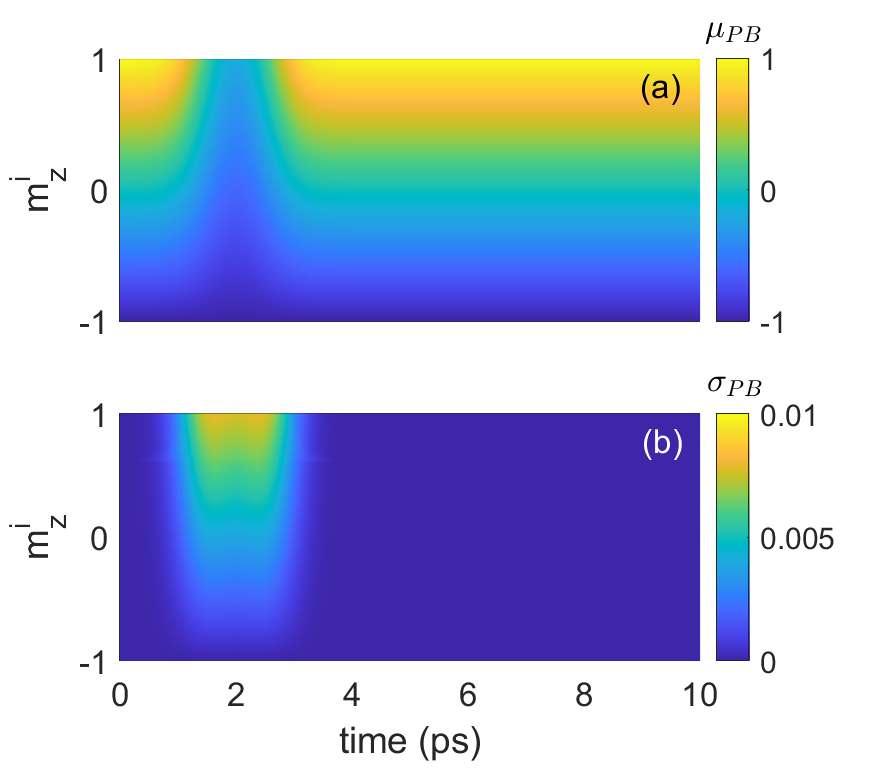}
\caption{\label{fig:mu_sigma} (a) Mean and (b) standard deviation of the non-equilibrium thermal field calculated from Eq.~\eqref{eq:meanstd} subject to an ultrafast pulse and assuming $N=125$. The mean tends towards negative values at the peak of the pulse at $2$~ps while it follows the value of the initial magnetization $m_z^i$ otherwise. The field's standard deviation also increases close to the optical pulse peak. }
\end{figure}
The switching probabilities can be estimated from an Arrhenius process assuming, for simplicity, that it depends only on $T(t)$. Thus,
\begin{equation}
\label{eq:PB_sim}
\begin{cases}P(t) &= 1-e^{-T(t)/T_c},\\B&=0,\end{cases}
\end{equation}
where $T_c=775$~K is the Curie temperature~\cite{Gorchon2016}. The resulting mean and standard deviation for the $\mathbf{H}_{n-e}$ are shown in Fig.~\ref{fig:mu_sigma} when the system is discretized in cubic cells of side $2$~nm, leading to $N=125$, and the laser pulse is set to $T_0=700$~K. The colorplots show the parameters calculated for the initial magnetization $m_z^i$ as a function of time. In panel (a), the mean of the non-equilibrium field simply follows the initial magnetization orientation except in the temporal range where the optical pulse is present. In that case, the mean favors a negative orientation for all initial magnetization values. The standard deviation similarly varies only when the optical pulse is active, shown in panel (b). The magnitude of the standard deviation is relative to the $m_z$ magnetization component and it is quite low. This is a consequence of the large number of spins in the cell. The standard deviation increases as the cell-size is smaller. Computing mean and standard deviations is a time-consuming endeavor because all possible combinations must be considered according to Eqs.~\eqref{eq:meanstd} and \eqref{eq:probnm}. Therefore, these statistics are precomputed for each discretization and used as look-up tables for the dynamic simulations.

The system is first allowed to thermalize by setting a thermal bath at $T=300$~K. This step follows the typical thermal field $\mathbf{H}_\mathrm{th}$ that is uniformly distributed over a sphere with magnitude given by the fluctuation-dissipation theorem~\cite{dAquino2006}. The uniform distribution is achieved by the Box-Muller algorithm~\cite{BoxMuller}. The thermalized magnetization is subsequently used as initial conditions for the ultrafast dynamic simulations.

\begin{figure}[t]
\includegraphics[width=3in, trim={0 0.9in 0 1.in}, clip]{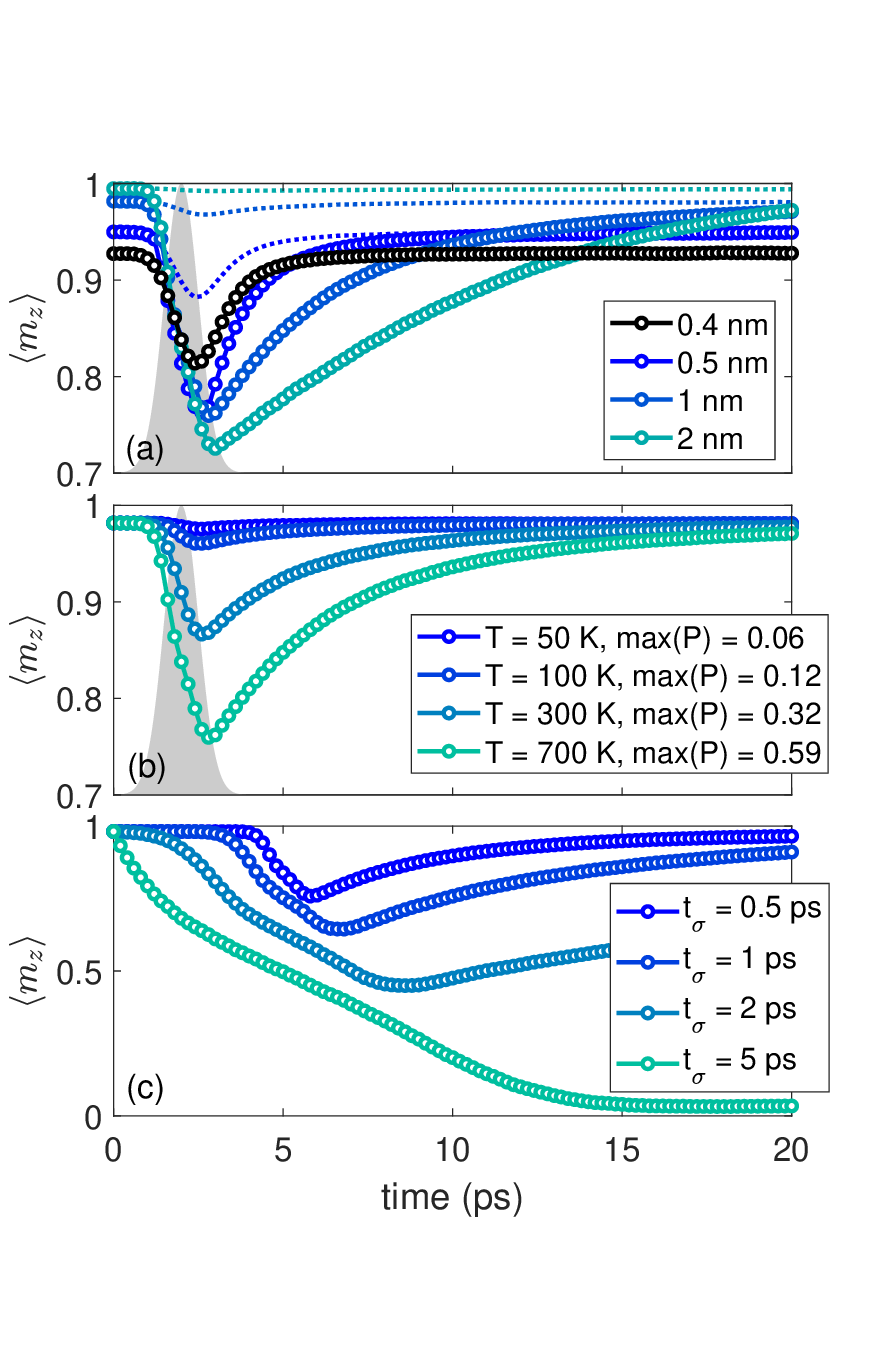}
\caption{\label{fig:results} Average $m_z$ magnetization component upon optical excitation for several cases. The laser pulse is shown in (a) and (b) as a guide to the eye. (a) Demagnetization as a function of cell size. The non-equilibrium field correction is zero in the atomic case, $a=0.4$~nm shown by black symbols. In that case, the demagnetization is purely driven by the uncorrelated noise. For the other cases, the demagnetization rates closely agree. (b) Demagnetization as a function of the maximum laser temperature. The equivalent maximum probabilities are shown in the legend. The increase in demagnetization is in qualitative agreement with experimental results. (c) Demagnetization as a function of the laser pulse standard deviation $t_\sigma$ and $T_0=700$~K. As the standard deviation increases, the magnetic system has more time to react to the non-equilibrium field and full demagnetization can be achieved.}
\end{figure}
In Fig.~\ref{fig:results}(a), we show the ultrafast evolution of the average $m_z$ magnetization component at several cubic cell discretizations. The optical pulse is shown as a gray area for reference. There are several salient features in the magnetization response. Before the optical pulses' peak, the simulations start at different averages. This is a direct consequence of the fluctuation-dissipation theorem approach where the larger cells are less susceptible to thermal fluctuations, on average. As the optical pulse acts on the magnetization, the thermal field is modified by the $\mathbf{H}_\mathrm{n-e}$ and all simulations exhibit a sharp decrease in the average $m_z$ component. Crucially, the decay rate is comparable for all discretizations, the only exception being $0.4$~nm which is the atomic resolution. This suggests that $|\mathbf{H}_\mathrm{n-e}|$ is likely overestimated by the presented approach. The basic assumption is that each spin flip occurs over a time interval $\delta t$, contributing to an energy of $\hbar/dt$. If the flip is incomplete in this time, $|\mathbf{H}_\mathrm{n-e}|$ will be reduced. 
It is also relevant to compare this response with a random thermal model. The ultrafast evolution of the average magnetization in each case is shown in Fig.~\ref{fig:results}(a) with dashed curves. We recognize the typical pathology of the fluctuation-dissipation theorem, i.e., a decreased impact of thermal fluctuations as the cell size increases. A particularly striking example is seen for a cubic cell of $2$~nm, where the purely thermal response is almost imperceptible while the addition of the non-equilibrium field drives demagnetization.

After the optical pulse passes, the evolution of the magnetization is completely determined by the internal energies, the thermal bath, and the boundary conditions. 
The different magnetization evolutions are thus expected based on the different magnon relaxation pathways and the evolution of solitons~\cite{Iacocca2019} that strongly depend on the discretization~\cite{Rockwell2024}.


Based on the above results, we can investigate the impact of the laser temperature and the spin switching probability, $P(t)$. In Fig.~\ref{fig:results}(b), we plot the average $m_z$ magnetization component for several laser peak temperatures and using cubic cells of $1$~nm. The temperatures and associated maximum switching probability $P_{nm}$ are listed in the legend. The quench is proportional to the temperature or maximum probability, as expected. It is also noteworthy that in all cases there is a delay between the maximum laser intensity and the minimum $\langle m_z\rangle$ which is found in the $1$~ps range. This value is consistent with the typical demagnetization timescales found in experiments~\cite{Kirilyuk2010}.

Finally, we explore complete demagnetization. In the results of Fig.~\ref{fig:results}(b), the demagnetization was limited to 0.7, even by increasing the laser temperature until a probability of 1 was achieved. The reason behind this failed demagnetization is the time needed to switch the magnetization in a cell and the exchange interaction counteracting this flip. For this reason, we explore the influence of the laser standard deviation at a $T_0=700$~K. The results are shown in Fig.~\ref{fig:results}(c) for several $t_\sigma$. We also shifted $t_c=5$~ps to model the growth and decay of the pulse. Clearly, as the laser acts on the magnetization for a longer time, the magnetization is quenched until full demagnetization is obtained for $t_\sigma=5$~ps. Full demagnetization corresponds to the formation of a labyrinth domain pattern for our modeled PMA material.

In conclusion, we presented a formulation for a non-equilibrium thermal field due to a femtosecond laser pulse with a defined helicity and whose magnitude depends on an effective magnon energy. The main assumption is that the femtosecond laser induces atomic spin flips which are modeled as a switching probability. In a cell consisting of many atomic spins, these probabilities define a mean field direction and standard deviation. 
While we numerically analyzed only the case of magnetization reversal of a ferromagnet, the approach takes into account both helicities. An extension to ferrimagnets is also possible insofar as these materials can be represented micromagnetically~\cite{Oezelt2015} or within a pseudospectral approach. In addition, the temporal dependence of magnetic parameters based on quasi-equilibrium temperatures, i.e., a Landau-Lifshitz-Bloch model~\cite{Raposo2022}, could be incorporated. Further refinements of this  approach can be achieved by considering incomplete switching, i.e., scaling the magnon energy by an effective spin flip time. It would be also interesting to implement this approach for larger areas and investigate the switching of ferromagnet due to multiple optical pulses~\cite{Mangin2014}.

The presented non-equilibrium field is able to achieve demagnetization curves within a micromagnetic model comparable to that obtained with atomistic spin dynamics, suggesting a method towards multiscale far-from-equilibrium dynamics and the numerical investigation of simulations comparable with experimental sample dimensions and laser diameters.

\section*{Acknowledgments}

This work was supported by the U.S. Department of Energy, Office of Basic Energy Sciences under Award Number DE-SC0024339.

\section*{Data Availability}

The data that support the findings of this article are openly available~\cite{OSF}.

\end{document}